\documentclass{aastex}

\newcommand{\Rmnum}[1]{\expandafter\@slowromancap\romannumeral #1@}

\begin{document}
\title{Recent glitches detected in the Crab pulsar}
\shorttitle{Glitches in the Crab pulsar}
\shortauthors{Jingbo Wang et al.}
\author{Jingbo Wang\altaffilmark{1,2}}
\and \author{Na Wang\altaffilmark{1,3}}
\and \author{Hao Tong\altaffilmark{4}}
\and \author{Jianping Yuan\altaffilmark{1}}
\altaffiltext{1}{Xinjiang Astronomical Observatory, Chinese Academy of Science,
 40-5 South Beijing Road, Urumqi, Xinjiang, China, 830011}
\altaffiltext{2}{Graduate School, Chinese Academy of Sciences, Beijing, China, 100049}
\altaffiltext{3}{Key Laboratory of Radio Astronomy, Chinese Academy of Science, Nanjing, China, 210008; na.wang@uao.ac.cn}
\altaffiltext{4}{Institute of High Energy Physics, Chinese Academy of Sciences, Beijing, China, 100049}
\begin{abstract}
From 2000 to 2010, monitoring of radio emission from the Crab pulsar at Xinjiang Observatory detected a total
of nine glitches. The occurrence of glitches appears to be a random process as described by
 previous researches.
A persistent change in pulse frequency and pulse frequency derivative after each glitch was
found. There is no obvious correlation between glitch sizes and the time since last glitch.
For these glitches $\Delta\nu_{p}$ and $\Delta\dot{\nu}_{p}$ span two orders of magnitude.
The pulsar suffered the
largest frequency jump ever seen on MJD 53067.1. The size of the
glitch is $\sim$ 6.8 $\times 10^{-6}$ Hz, $\sim$ 3.5 times that of the glitch occured in 1989
glitch, with a very large permanent changes in frequency and
pulse frequency derivative and followed by a decay with time constant $\sim$ 21
days. The braking index presents significant changes. We
attribute this variation to a varying particle wind strength which
may be caused by glitch activities. We discuss the properties of
 detected glitches in Crab pulsar and compare them with
glitches in the Vela pulsar.
\end{abstract}


\section{Introduction}
\label{sec:intro}
The Crab Nebula is the remnant of supernova explosion recorded by Chinese astronomers in AD 1054.
It is the archetype of center-filled supernova remnants (or plerions) with a distant of $\sim$2 kpc and one
of most well-studied objects in almost all wavebands from low frequency radio to very high energy
$\gamma$-ray. The overall non-thermal radiation of Crab nebula is mainly dominated by synchrotron process
and inverse compton scattering. The synchrotron origin of the optical and radio continuum emission
was proposed by Shklovskii (1953) and experimentally confirmed by polarization observations (Dombrovsky 1954).

The nebula is powered by the central Crab pulsar (PSR B0531+21).
It is the second brightest pulsar in the northern sky at radio waveband and visible through the whole
observable electromagnetic spectrum. In addition, the pulsar is one of the best studied and most energetic
pulsar. The spin down power for the pulsar is $\dot{E} = 4.6 \times 10^{38}$ $\mathrm{erg}$ $s^{-1}$. Approximately
 10\% - 20\% of the spin-down energy of the pulsar is converted into the radiation of the nebula. Its pulse
frequency ($\nu$ = 30 Hz) and pulse frequency derivative ($\dot\nu= -3.7 \times 10^{-10}$ $\mathrm{s^{-2}}$)
yields an characteristic
age of 1240 years, very close to the actual age.

The average emission profiles of Crab pulsar are dominated by a main pulse (MP) and an interpulse (IP).
The MP and IP are separated $\sim$ 0.4 in pulse phase in all wavelength bands. But the pulse peaks are not
fully aligned in phase over the entire energy range. Previous timing of the pulsar over many orders of
magnitude of energy found that there is a delay of the radio main pulse compared to the first peak seen
from optical to hard $\gamma$-rays. Recent measurements of the optical, X, $\gamma$-ray to radio lag are:
255 $\mu$s (Oosterbroek et al. 2008), 275 $\mu$s (S. Molkov et al. 2009), 281 $\mu$s (Abdo et al. 2010),
 respectively. The small differences in pulse phase alignment allow
us to study the emission regions at very small scales.
Romani \& Yadigaroglu (1995) have proposed that the radio precursor comes at the polar cap,
whereas the main pulse and interpulse originate in the outer gap of the magnetosphere, and
higher energy pulses are generated at significantly greater heights.
Therefore, precise timing of mean pulse
profiles over the full range of electromagnetic spectrum is a powerful tool for understanding of
the nature and spatial origin of the emission mechanisms.

The Crab Pulsar, just like many young pulsars, is influenced by significant glitches and timing noise in pulse frequency.
Timing noise in the Crab pulsar is the dominant component in the timing residuals after removal of the spin-down
model and glitch effects from the pulse phase and seen as continuous, noise-like fluctuation
in frequency.
Scott et al. (2003) shows the timing noise is composed of two
components: a long-term quasi-periodic oscillation with
a period of 568 $\pm$ 10 days and a red noise process with an
approximately \textit{f}$^{-3}$ power density spectrum.

 Pulsar glitches are characterized by the sudden increases in pulse frequency.
Glitches are rather rare and unpredictable phenomena, and
they vary significantly for different pulsars.
The characteristics of glitches and the post-glitch recovery
behavior provide an important diagnostic tool to study neutron star interiors (e.g., Ruderman
1969; Baym et al. 1969). For the Crab pulsar, twenty four glitches have been detected from 1969 to 2008 (Lyne et al.
1993; Wong et al. 2001; Wang et al. 2001a; Espinoza et al. 2011a). The glitches of Crab pulsar
are characterized by their small relative size, rapid exponential relaxation
towards extrapolated pre-glitch frequency and a persistent change in frequency derivative at each
glitch. Moreover, the large glitches in 1989 and 1996 manifests gradual spin-up right after the initial
frequency jump and each glitch may accompanied by an "aftershock" or secondary spin-ups
20-40 days after an event. The persistent increase in the magnitude of pulse frequency
derivative may be due either to an increase in the external torque (Link et al. 1992) or
a variation in the momentum of inertia acted on by the torque (Alpar et al. 1996).

 In order to study the glitch process, long-term monitoring of frequent glitching pulsars such as Crab and Vela are
necessary. Here we present timing observations of the Crab pulsar from 2000 to 2010. Nine glitches have been
observed at Xinjiang Astronomical Observatory during this period.
\section{Observations and Data Reduction}
Timing observations of Crab pulsar at 1540 MHz commenced in 2000 January as part of pulsar monitoring
program at Xinjiang Astronomical Observatory (Wang et al. 2001b), with about one observing session per week.
A dual-channel room temperature receiver ($\sim$ 100 K) was used
and then updated to a cryogenic system in 2002 July, which allow us to detect pulsars with a mean flux
density greater than 0.5 mJy. The two hands of circular polarization are sent to a
filter-bank consisting of 2$\times$128 channels of width 2.5 MHz. The data are 1-bit digitized and
sampled at 1-ms intervals. Time is provided by a hydrogen maser calibrated by the Global Position System
(GPS) and a latched microsecond counter. The integration time of each observation for the Crab pulsar is 16 minutes.

The offline data was dedispersed to remove the dispersion effects of interstellar medium and
folded at nominal topocentric period to produce a mean pulse
profile. Topocentric times of arrival (ToAs) were obtained by
cross-correlating the mean pulse profile with a noise-free template
using {\sc psrchive} soft package (Hotan et al. 2004).
Local arrival times were converted to Solar-system barycenter times
by using {\sc tempo2}\footnote{See http://www.atnf.csiro.au/research/pulsar/tempo2/.}(Hobbs et al. 2006) with Jet
Propulsion Laboratory ephemeris DE405 (Standish 1998). ToAs are weighted by the inverse square of their uncertainty.
Uncertainties in the fitted parameters are taken to be twice the formal uncertainties obtained from {\sc tempo2}.

The pulse phase $\phi$ predicted by standard timing model is expressed as:
\begin{equation}
   \phi(t)=\phi_{0}+\nu(t-t_{0})+\frac{1}{2}\dot\nu(t-t_{0})^{2}+
         \frac{1}{6}\ddot\nu(t-t_{0})^{3},
\end{equation}
where $\phi_{0}$ is the pulse phase at reference time $t_{0}$.

 Glitches are usually described as combinations of
step changes of $\nu$ and $\dot\nu$, $\ddot\nu$, parts of which decay exponentially:
\begin{equation}
 \nu(t)=\nu_{0}(t) + \Delta\nu_{p} + \Delta\dot{\nu}_{p}t + \frac{1}{2}\Delta\ddot{\nu}_{p}t^{2} +
 \Delta\nu_{d}\;e^{-t/\tau_{d}},
\end{equation}
\begin{equation}
\dot{\nu}(t)=\dot{\nu}_0(t) + \Delta\dot{\nu}_p + \Delta\ddot{\nu}_pt +
\Delta\dot\nu_{d}\;e^{-t/\tau_{d}},
\end{equation}
where $\nu_{0}(t)$ and $\dot\nu_{0}(t)$ are the pulse frequency and pulse frequency derivative extrapolated from
pre-glitch model, respectively; $\Delta{\nu}_p$, $\Delta\dot{\nu}_p$ and $\Delta\ddot{\nu}_p$
are the permanent
changes in frequency, its first and second derivatives relative to the extrapolated pre-glitch values,
respectively, and $\Delta\nu_{d}$ is the amplitude of an exponential relaxation component with
a decay time constant of ${\tau}_d$. The total frequency and frequency derivative changes at
the time of the glitch are $\Delta\nu_{g} = \Delta\nu_{d} + \Delta\nu_{p}$ and
$\Delta\dot{\nu}_{g} = \Delta\dot{\nu}_{d} + \Delta\dot{\nu}_{p}$, respectively.
And the degree of recovery can be described by parameter: $Q = \Delta\nu_d/\Delta \nu_{g}$.
The pulsar position used in the reduction is
taken from the Jodrell Bank Crab Pulsar Timing Results Monthly Ephemeris\footnote{See http://www.jb.man.ac.uk/pulsar/crab.html.}(Lyne et al. 1993).
\section{Results}
Nine glitches have been observed in the Crab pulsar during the period from 1999 October to
2010 September (MJD 51455 to MJD 55446).
However, according to Espinoza et al.(2011a), there were eleven glitches during the period.
We missed two small glitches (5 and 9) because of the observation gaps.
Table 1 lists the pre- and post-glitch rotation parameters.
Estimated uncertainties in the last quoted digit
are given in brackets. As shown in Table 1, several interglitch intervals are very short.
It is quite difficult to get the value of $\ddot\nu$ for such a short data span. In addition,
these $\ddot\nu$ values could be affected by the effect of post-glitch exponential decay.
Therefore, we keep the $\ddot\nu$ at a fixed value which can make the braking index equal to 2.51 for these very short interglitch intervals.
The glitch parameters are given in Table 2.
The glitch parameters except the glitch epoch were obtained from {\sc tempo2}.
The glitch epoches given in the second column of Table 2 which have accuracy of $\sim$ 0.1 day
were obtained from the Ephemeris (Lyne et al. 1993).
We list the two missed glitches (5 and 9) in Table 2
for completeness with the glitch sizes taken from Espinoza et al.(2011a).
Because of the short interglitch intervals and observation gaps,
our observations around the epoch of most glitches are not frequent enough to determine
the decay timescale.
In general, our results are consistent with Espinoza et al. (2011a). Instead of directly fitting glitch
parameters in {\sc tempo2}, the glitch parameters
given by Espinoza et al. (2011a) were obtained by comparing the timing solutions before and after the glitch.
\begin{table*}
\tiny
 \caption{Pre- and post-glitch timing solutions.}

 \begin{tabular}{lllllllll}
\tableline
Int.&$\nu$                   &$\dot{\nu}$           &$\ddot{\nu}$~~&Epoch   &MJD Range &No. of &RmsRes\\
    & (s$^{-1}$)&($10^{-9}$~s$^{-2})$  & $(10^{-21}$~s$^{-3})$ &(MJD)         &           &ToAs  &($\mu$s) \\
\tableline
$-$1   & 29.843629669(3)  & $-$0.3745060(4) &10.8(3) &51622 &51547$-$51738& 33&1399  \\
1$-$2  & 29.838744465(6) & $-$0.3744321(79)&11.8&51773 &51745$-$52080& 11&803  \\
2$-$3  & 29.833181422(14) & $-$0.3742666(2)&9.4(1) &51945 &51824$-$52080&36 &723  \\
3$-$4  & 29.827781949(6) & $-$0.3742107(81) &11.8 &52112 &52088$-$52140&10  &830  \\
4$-$5  & 29.820962244(1) & $-$0.3798937(1)&10.7(6)&52323 &51151$-$52495&40 &987  \\
5$-$6  &29.813984966(4) & $-$0.3738076(43)&11.7 &52539 &52503$-$52574&11 &685  \\
6$-$7  & 29.803976844(1) & $-$0.3735181(1)&10.8(3)&52849 &52605$-$53063&63 &1085  \\
7$-$8  &29.792879069(4)  & $-$0.3734151(8) &19.9(11)  &53193 &53133$-$53253&36 &698  \\
8$-$9  &29.789846906(3)  & $-$0.3733023(30) &11.7  &53287 &53260$-$53315&8 &363  \\
9$-$10 &29.7725067846(7)  & $-$0.37277839(1)&11.45(4)&53825 &53685$-$53965&161&814  \\
10$-$11&29.752746656(1)  &$-$0.3721811(1)&11.7(6)&54439 &54306$-$54566&95&700  \\
11$-$   &29.7282686152(5)  &$-$0.37141433(4)  &11.93(1)&55201 &54949$-$55194&194&1150\\
\tableline
\end{tabular}
\end{table*}

The rotation history is
presented in Fig.1 and Fig.2. The values of frequency and frequency derivative in the figures were
obtained from independent fits to short sections of the data. The uncertainties of
frequency and frequency derivative in the
plots are 1$\sigma$ from the {\sc tempo2} fit. These figures include timing residuals in
phase relative to the pre-glitch timing model (panel (a)), variation of frequencies obtained at different
epochs relative to the pre-glitch model (panel (b)), and variations of frequency derivatives (panel (c)).
 These subplots in each figure are arranged in chronological order.
We discuss the detailed results of the nine glitches in the following sections.
\begin{table*} \tiny
\caption{The glitch parameters.}
\begin{tabular}{lllllllllll}
\tableline
GLT.&Glitch Epoch&$\Delta\nu_g$& $\Delta\nu_p$&$\Delta\dot\nu_p$& Q    &$\tau_{d}$&MJD Range&RmsRes\\
No. &MJD(date)   &($10^{-6}$~$s^{-1})$&($10^{-6}$~$s^{-1}$)&($10^{-15}$~$s^{-2}$)&&$(d)$&&($\mu$s)\\
 \tableline
1 &51739.4(000714) &---     &0.174(32) &$-$54(10)  &---&---&51505$-$51800&1089 \\
2 &51804.9(000917) &---     &0.029(21) &$-$6(10)   &---&---&51745$-$51938&608 \\
3 &52083.8(000624) &---     &0.409(17) &$-$79(5)   &---&---&51811$-$52140&792\\
4 &52146.0(010824) &---     &0.132(23) &$-$3(9)    &---&---&52088$-$52301&607\\
5 &52497.3(020812) &0.101(3)$^{a}$&---&---         &---&---&---          &---\\
6 &52587.1(021109) &---     &0.046(23) &$-$5(8)    &---&---&52503$-$52772&788\\
7 &53067.1(040303) &6.76(13)&1.211(56) &$-$202(5)  &0.82(2)&21.1(8)&52557$-$53216&1038\\
8 &53254.2(040910) &---     &0.090(21) &$-$17(22)  &---&---&53166$-$53320&598 \\
9 &53331.1(041122) &0.08(1)$^{a}$&---       &---   &---&---&---       &---\\
10 &53970.0(060822)&0.41(9) &0.132(6)&$-$19(1)&0.68(8)&7.3(34)&53712$-$54142&761 \\
11 &54580.0(080423) &---    &0.516(89)  &$-$7(6)   &--- &---&54450$-$54672&579\\
\tableline
\end{tabular}
\tablenotetext{a}{The glitch sizes are taken from Espinoza et al. (2011a).}
\end{table*}

\subsection{The largest glitch}
The Crab pulsar suffered its largest frequency jump ($\Delta\nu_{g}/\nu \sim 2\times10^{-7}$)
ever seen in 2004 March (MJD 53067.1). The size of the
event is more than 3.5 times that of the glitch occured in 1989. Despite the fraction of increase in frequency is still
much smaller than the typical value of the Vela pulsar, the glitch amplitude is comparable to small glitches
of the Vela pulsar. Although we could not identify the exponential decay process from Figure 1,
we can still obtaine the decay time constant and the amplitude of the transient frequency jump
with high significance by fitting the two parameters in {\sc tempo2}.
 The recovery fraction is $\sim$ 82\% and the decay
time constant is $\sim$ 21 days which is largest among all the glitches in the Crab pulsar.
The glitches occured in 1969, 1975 and 1989 have a smaller decay time constant than Glitch 7
(Lyne et al. 1993; Wong et al. 2001). As shown in Table 2 and Figure 1, there was a remarkable
permanent increase in pulse frequency and the magnitude of the frequency derivative, $\Delta\nu_{p} \sim 1.2\times10^{-6}$ Hz
and $\Delta\dot\nu_{p} \sim -202\times10^{-15}s^{-2}$, respectively. Note that these two parameters are also
larger than that of any other glitch in the Crab pulsar.
\begin{figure*}[htbp]
\centering
\includegraphics[height=9cm, angle=270]{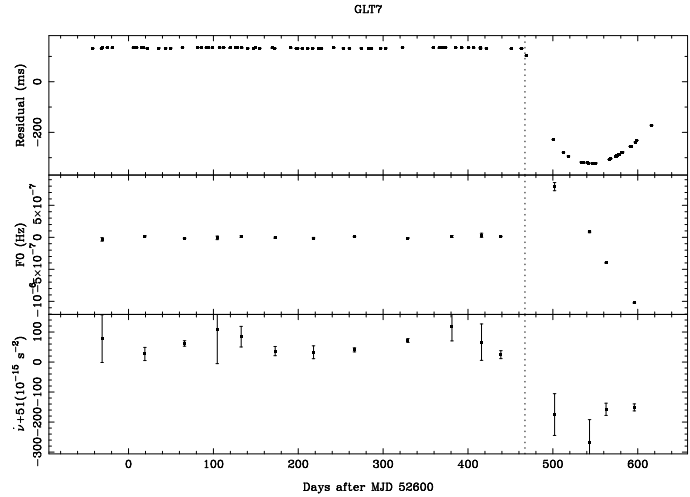}
\caption{The largest glitch (Glitch 7 of this work) : (a) timing residuals relative to the pre-glitch model. (b) variations
of the frequency residuals $\Delta\nu$ relative to the pre-glitch solution. (c) the
variations of $\dot\nu$. The dashed vertical line indicates the epoch of the glitch.}
\end{figure*}
\subsection{The other glitches}
The frequency jump of Glitch 1 can be easily recognized from the middle panel of GLT1 in Fig.2.
The post-glitch increment of pulse frequency derivative can be seen in the bottom panel as well,
and this glitch caused a permanent change of pulse frequency derivative. Glitch 2 is a
much smaller event and only a little frequency jump could be seen in GLT2 of Fig.2.
As shown in Table 2, no evident $\Delta\nu_{p}$ is measured.
The subplot GLT3 in Fig.2 is very similar to that of GLT1. However, the amplitude of $\Delta\nu_{p}$ and
$\Delta\dot\nu_{p}$ are larger than GLT1. Glitch 4 is another small glitch. Compared with the pre-glitch
solution, we can see the pulse frequency after the glitch
decreased gradually from the frequency residuals plot of Glitch 4. But the pulse frequency derivative of Glitch 4
almost remain the same for about 100 days after the glitch and decreased then. And Glitch 6 is a small glitch
as well. The timing and frequency
residuals plots of Glitch 6 look similar to that of Glitch 4. The pulse frequency derivative
after the glitch exhibit a small fluctuation and no obvious $\Delta\dot\nu_{p}$ is detected as well.
Apparently, as seen from GLT8 of Figure 2, we miss the frequency jump of Glitch 8
because of the observation gap. However, as shown in Table 2, we got a significant value for $\Delta\nu_{p}$
and the $\Delta\dot\nu_{p}$ induced by the glitch can be identified from the bottom panel of GLT8.
Glitch 10 is a relative large one in magnitude with a small recovery fraction $\sim$ 68\% and a short decay time constant
. The exponential fit for the glitch in 1996 (Wong et al.
1996) has nearly the same recovery fraction, but the measured $\Delta\nu_{p}$, $\Delta\dot\nu_{p}$, and
decay time constant of Glitch 10 are much smaller than that of the glitch in 1996. The last glitch of
this paper occured on MJD 54580.0 with a $\Delta\nu_{p}$ value similar to Glitch 6. In contrast to
pre-glitch timing model, The post frequency residuals decreased steady which indicate a change in the
pulse frequency derivative and this could be also seen in the lower panel of GLT11 in Fig.2.
\begin{figure*}[htbp]
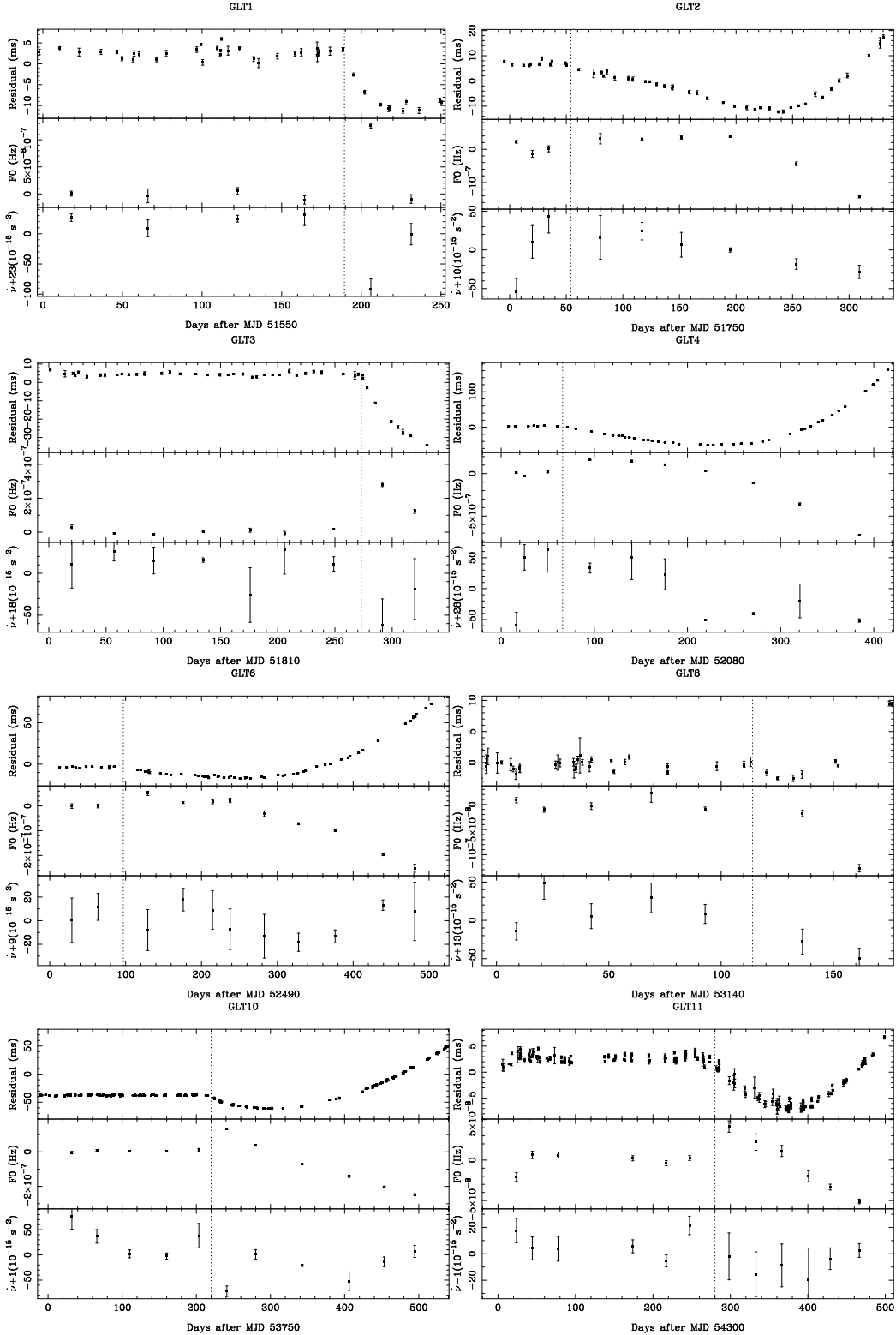

\includegraphics[height=6.9cm, angle=270]{glt1.ps}%
\includegraphics[height=6.9cm, angle=270]{glt2.ps}
\includegraphics[height=6.9cm, angle=270]{glt3.ps}%
\includegraphics[height=6.9cm, angle=270]{glt4.ps}
\includegraphics[height=6.9cm, angle=270]{glt6.ps}%
\includegraphics[height=6.9cm, angle=270]{glt8.ps}
\includegraphics[height=6.9cm, angle=270]{glt10.ps}%
\includegraphics[height=6.9cm, angle=270]{glt11.ps}
\caption{Eight glitches (Glitch 1, 2, 3, 4, 6, 8, 10 and 11 of this work) : (a)
timing residuals relative to the pre-glitch model. (b) variations
of the frequency residuals $\Delta\nu$ relative to the pre-glitch solution. (c) the
variations of $\dot\nu$. The dashed vertical line indicates the epoch of the glitch.}
\end{figure*}
\section{Discussion}
\subsection{The interglitch intervals}
Wong et al. (2001) pointed out that the Crab pulsar glitches appear to be
independent events spaced randomly in time. However there were only eight
glitches and seven interglitch interval by the time of their work.
 Therefore, we add the eleven new glitches to the sample for the statistics,
which consists a total sample of 19 glitches.

The left panel of Fig.3 shows the observed distribution of interglitch intervals of Crab pulsar from 1983 to
2008. The dashed line represents expected distribution for a Poisson process with average interval $\lambda$,
\begin{equation}
P(T) = 1 - e^{-t/\lambda},
\end{equation}
where $\lambda$ is the mean interglitch interval. The plot shows that for the Crab pulsar
the occurrence times of the glitches is well described by Poisson distribution.
Fitting of the data gives a mean interglitch
interval of 419 days which is significantly smaller than
the previous result of 684 days given by Wong et al. (2001), the frequent glitches in recent
years mainly contribute to the shorter interval. The standard deviation of
the interglitch interval is quite large, 365 days.
A Kolmogorov-Smirnov (K-S) test obtained a probability P$_{K-S}$ = 0.991,
This means that the distribution of interglitch interval agrees well with a Poisson model.
Melatos et al. (2008) also found that the occurrence times of the glitches are consistent with Possion
statistics in the Crab and several other glitching pulsars.

\begin{figure*}[htbp]
\includegraphics[height=6.0cm]{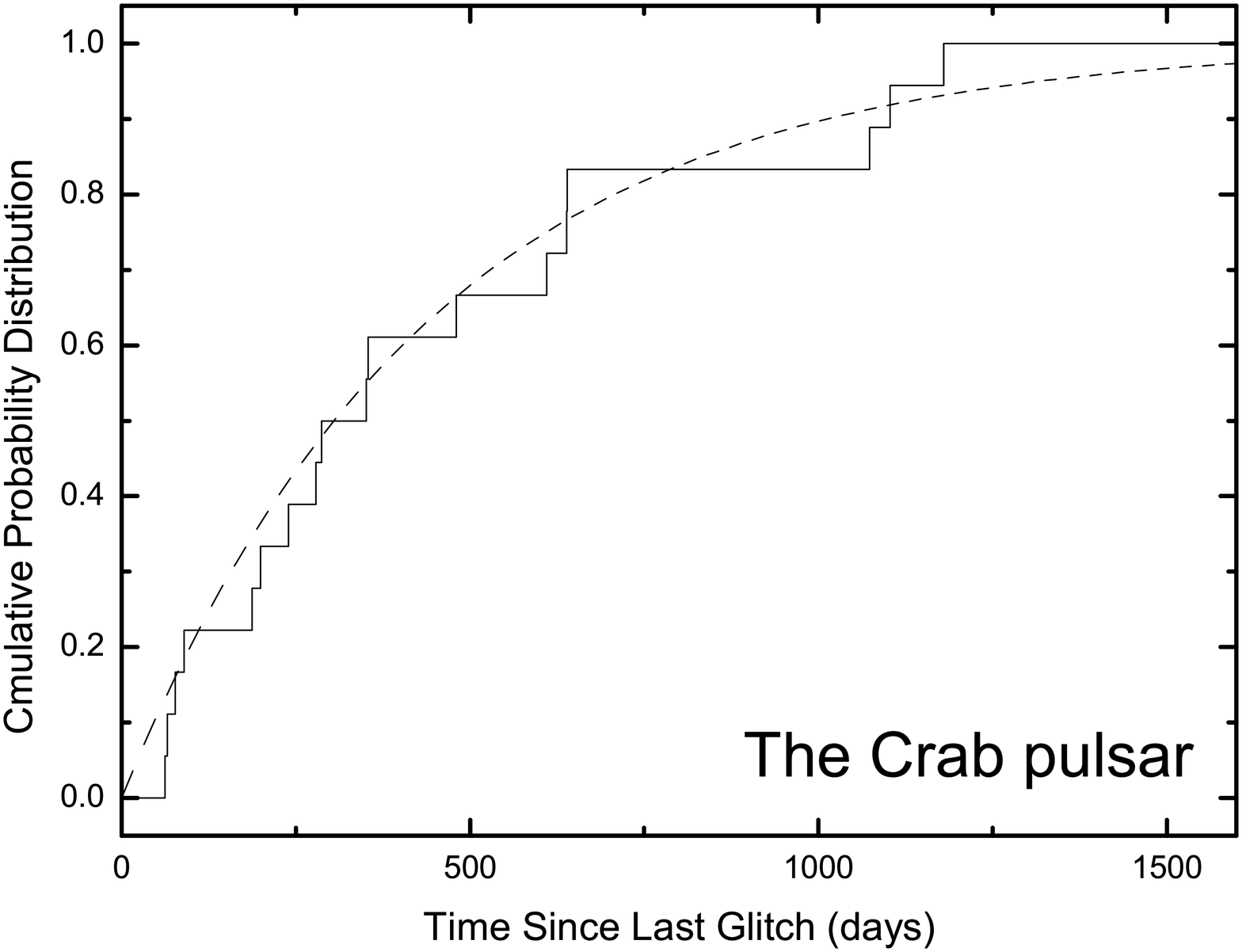}%
\includegraphics[height=6.0cm]{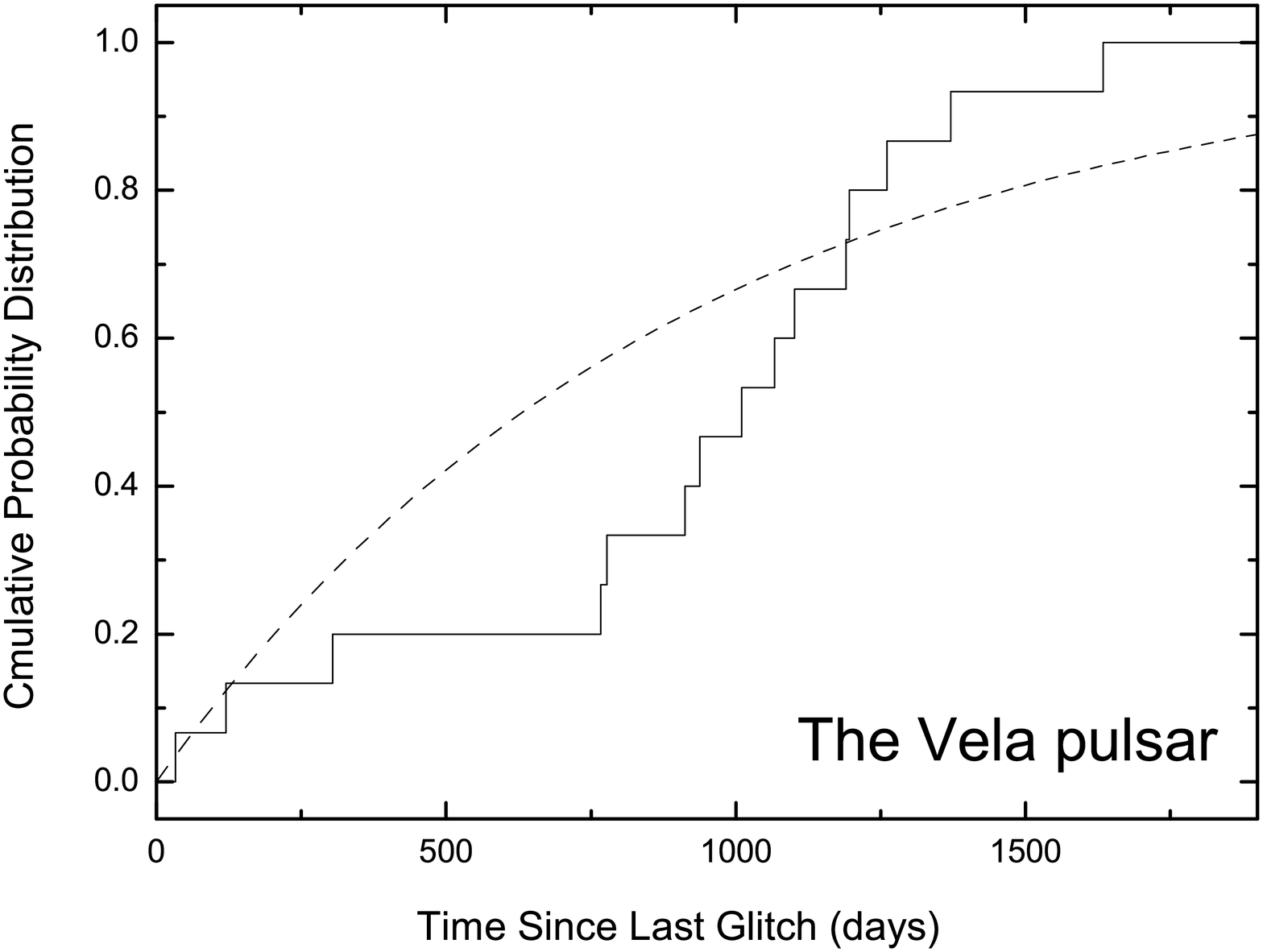}
\caption{Cumulative distribution of interglitch intervals for the Crab (left) and the Vela (right) pulsar.
The dashed curves represent a Poisson distribution with a mean interval of 419 and 912 days
for them, respectively.}
\end{figure*}
By comparison, we present the interglitch intervals of the Vela pulsar and Possion distribution in the right panel of Fig.3.
In total 16 glitches from 1969 to 2006 are included (Cordes et al. 1988, McCulloch et al.
1987, McCulloch et al. 1990, Flanagan 1991, Flanagan 1994a,
Flanagan 1994b, Wang et al. 2000, Dodson et al. 2002,
Dodson et al. 2004, Lanagan \& Buchner 2006).
This reveals a mean glitch interval of 912 days.
Fig.3 clearly shows that the distribution of interglitch interval
is essentially different from a Poisson prediction for the Vela pulsar.
A K-S test with $\lambda$ = 912 days shows that we can reject the Possion model at a 93.9\% confidence level.
Melatos et al. (2008) shows the glitch occurence is quasi-periodic.
The typical and relatively quasi-periodic glitch occurence present in each pulsar
might indicate that a critical lag between the rotation of the superfluid and crust has to be achieved in
order for a glitch to occur (Alpar et al. 1993). Figure 4 shows the relation between the fractional
glitch size and the time since previous glitch for the Crab and the Vela pulsars. The size of glitches
for the Crab pulsar are obtained from Espinoza et al. (2011a) and this work. The Crab pulsar has a wide
range in both parameters while most glitches of Vela pulsar are large and have longer interglitch intervals.
For both pulsars, little correlation are found between glitch sizes and the time since last glitch, even though
theoretical predictions prefer that the amplitude of a glitch would be proportional to the intervals of preceding glitch
 (e.g., Alpar et al. 1989; Ruderman et al. 1998).
\subsection{Properties of observed glitches}
Some pulsars like the Crab only experienced small glitches (e.g., Yuan et al. 2010; Krawczyk et al. 2003),
while Vela-like pulsar are characterized by large size glitches (e.g., Shemar \& Lyne 1996; Wang et al. 2000).
Some pulsars exhibit wide glitch size distribution spanning three or even four orders of magnitude ( 10$^{-10}$
to 10$^{-6}$ Hz; e.g., Lyne 1987; Janssen \& Stappers 2006). The absence of large glitch in the
Crab pulsar is likely attributed to its relative young age; the
stresses build up during steady spin down can partly relieved by gradual process such as vortex creep and
plastic flow at high crustal temperature (e.g., Ruderman 1991; McKenna \& Lyne 1990).
 As shown in Figure 4, there is no significant correlation between the glitch size and the time since last glitch
for the Crab and Vela pulsars.
 In addition, lack of correlation between glitch size and preceding interglitch
interval in most pulsars (e.g., Yuan et al. 2010; Wang et al. 2010) suggests that the triggering
of glitches is not closely related to global slowdown of the pulsar.

\begin{figure}[htbp]
\includegraphics[height=6cm]{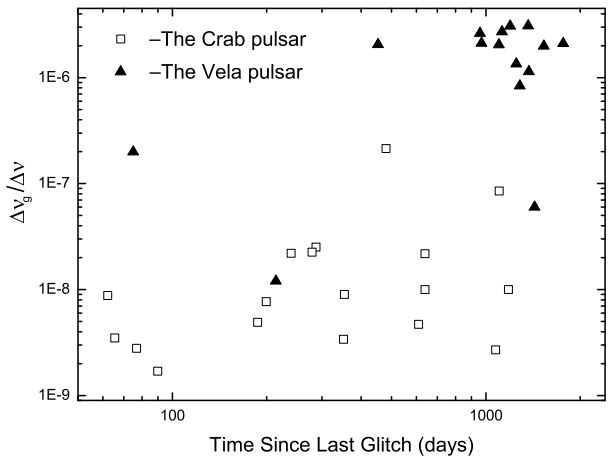}
\caption{Glitch size against the time since the
previous glitch since 1989 (earlier glitches have been omitted as a result of possible gaps in the
timing record). Squares and filled triangles indicate the Crab and the Vela pulsars glitches, respectively.}
\end{figure}

The activity parameter $A_{g} \equiv
(\Sigma\Delta\nu_{p})/t_{obs}$ is defined to be a long-term indicator of
glitch effects (Wong et al. 2001). Figure 5 shows a cumulative plot
of persistent changes in $\nu$ as a function of time for the Crab pulsar since 1969. The
activity parameter $A_{g}$, which is represented by the slope of
this relation, is about 1.4 $\times 10^{-5}|\dot\nu| $. This value is
essentially the same as that of given by Wong et al. (2001). It implies
that the rate of angular momentum loss caused by glitch has not
changed significantly, in spite of high glitch rate during the past 10 yr.
\begin{figure}[htbp]
\includegraphics[height=6cm]{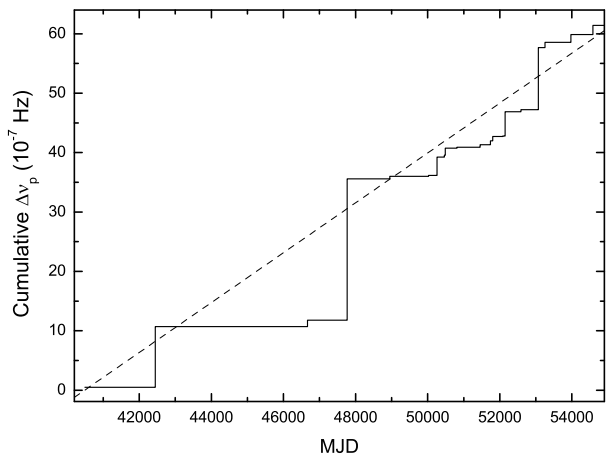}
\caption{Cumulative $\Delta\nu_{p}$ caused by glitches plotted as a function of time based on data
from Table 2. The dashed line represent a least-squares fit to the midpoints of the frequency jumps . The
slope of the dashed line is an estimate of the activity parameter A$_{g}$.}
\end{figure}
The persistent changes in $\dot\nu$ caused by the Crab pulsar glitches is probably due to the formation
of new vortex capacitors (Alpar et al. 1996). Superfluid decouples from the steady slow down of star crust
leading to the decrease of moment of inertia. Therefore, the external torque acts on a lower moment of
inertia and thus $|\dot\nu|$ increases. On the other hand, the variation of external torque result
in change in $\dot\nu$, maybe due to an increase in the dipole magnetic field (Ruderman et al. 1998) or a
change angle between magnetic and rotation axis (Link et al. 1998).
A long term asymptotic exponential rise in frequency, as shown in
Figure 7b of Lyne et al. (1993) was observed after the 1975, 1981, 1986 and 1989
glitches. However, no similar event has been observed since.
This is probably because the timescale of asymptotic rise exceeds the interglitch intervals.
The wide range of glitch parameters implies a local phenomenon of the
Crab pulsar in which starquakes could be responsible for Crab glitches.
The small size of glitches, permanent postglitch offsets in the frequency derivative and the
large recovery fraction of Crab pulsar glitches shown by our and previous results are all consistent
with starquakes model (e.g., Lyne 1992; Alpar et al. 1994, 1996; Link et al. 1998; Franco et al. 2000;
 Crawford and Demianski 2003). By contrast, starquakes can not be
the cause of glitches for the Vela pulsar (Lyne 1992; Alpar et al. 1993; Chau et al. 1993;
Alpar et al. 1995; Crawford and Demianski 2003).

\subsection{The braking index}
The high glitch rate makes it difficult to measure the braking index of the Crab pulsar.
 The interval between Glitch 9 and Glitch 10 is relatively longer.
And the Crab pulsar has not experienced any glitch after 2008 April (Glitch 11).
We measure the braking indices for these two data spans. Observed values of frequency
and its first and second derivates can be used to obtain the braking index \textit{n}, using the equation:
\begin{equation}
n = \nu\ddot\nu/\dot\nu^{2}.
\end{equation}
In order to avoid the effect of exponential decay of the steps in $\dot\nu$ and $\ddot\nu$
of the glitch, we omitted the observations about one year after the previous glitch.
The braking indices are calculated based on the timing parameters listed in Table 1 for these two data spans,
giving values of 2.454(7) and 2.571(3), respectively.
The previous measured value of braking index show a remarkable constant
value 2.51 (Lyne et al. 1993, Wong et al. 2001). It is clear that there is an evident change in braking index.
The $\Delta\ddot\nu$ changes have been found by Wong et al. (2001).
In fact, the changes in braking index
which have become more marked during the last 20 years or so, when the amount of glitch activity increased.
The value of braking index after Glitch 11 is larger than the previous measured value, whereas the other braking
index measured between Glitch 9 to Glitch 10 is much smaller. So the scatter of braking index is much larger
than that given by Lyne et al. (1993) and Wong et al. (2001).

The reason of a varying braking index may be due to a
varying particle wind. A particle wind in addition to the magnetic dipole
radiation may account for a braking index less than three
(Michel 1969; Manchester et al. 1985; Xu \& Qiao 2001; Espinoza et al. 2011b).
The existence of particle wind is verified by observations of
intermittent pulsars (Kramer et al. 2006; Camilo et al. 2011).
A fluctuating wind may contribute to the long term timing noise
(Lyne et al. 2010; Liu et al. 2011).
Another consequence of a varying particle wind
will be a varying braking index. This may be the
case of PSR J1846-0258 (Livingstone et al. 2011).
In this paper, we apply the pulsar wind model of Xu \& Qiao (2001) to the Crab pulsar.
(Other pulsar wind models are similar. They contain a dipole component
and wind component, e.g. Spitkovsky 2006.)
Employing the polar cap model of Ruderman \&
Sutherland (1975) and considering that the wind strength may depart
from the long term average value, Figure 6 shows the braking index
as a function of wind strength (Xu \& Qiao 2001)\footnote{For
discussions of different particle acceleration models see Xu \& Qiao
(2001). The corresponding calculations will only differ
quantitatively when employing different particle acceleration
models.}. A larger wind strength will cause the braking index
smaller. A wind strength of 1.19 will give a braking index 2.45, and
a wind strength of 0.87 corresponds to a braking index of 2.57.
Therefore, a varying wind strength will result in a varying braking
index naturally. The variations from long term average value are
mainly fluctuations. This fluctuations may be caused by frequent
glitches of the Crab pulsar, similar to the glitch induced magnetospheric activities
seen in magnetars (Kaspi et al. 2003).

Since we only have two braking index measurements at
present, an estimation of possible short term net increase in wind
strength is very uncertain. Based on these two measurements, the net
increase in wind strength is three percent. The corresponding relative change in
slow-down rate is 0.007. Accurate to one order of
magnitude, this is consistent with the largest increase in slow-down
rate after glitches as shown in Table 3.
\begin{figure}
  \includegraphics[height=5cm]{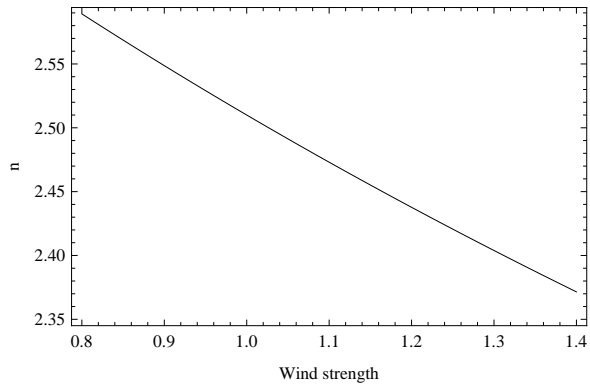}
  \caption{The braking index as a function of wind strength.
  Calculations for the Crab pulsar.
  Wind strength 1 is the long term average value, which corresponds to
  a braking index of 2.51 (Xu \& Qiao 2001). Wind strength 1.2 means the
  wind luminosity is 1.2 times the long term average value, etc.}
\end{figure}
\section{Summary}
We have presented timing observations of the Crab pulsar from 2000
to 2010. During this period, this pulsar manifest a higher rate of
glitch activities than previous, with nine glitches detected over a period of 8 yr.
The number of glitches in the Crab pulsar has increased considerably.
The distribution of interglitch intervals is still in agreement with
a random process.
There is no correlation between the glitch amplitude and the time
since the last glitch.
In accord with previous study (Lyne et al. 1993, Wong et al. 2001), permanent changes in
pulse frequency derivative are observed at the time of each glitch.
Since the relative small permanent frequency changes for the recent
glitches, the average pulse frequency derivative caused by glitches (represent by
the activity parameter $A_{g}$) almost remains the same, in spite of
the high glitching rate. The braking index shows an relative
obvious variation, which may be due to a varying particle wind
strength induced by glitches.

\section{Acknowledgments}
This work is supported by NSFC project 10673021, the Knowledge Innovation Program of the Chinese Academy of Sciences,
Grant No. KJCX2-YW-T09, National Basic Research Program of China (973 Program 2009CB824800) and West
 Light Foundation of CAS (No. XBBS201021).


\end{document}